\documentclass[fleqn,twoside]{article}
\usepackage[headings]{espcrc2}

\readRCS
$Id: espcrc2.tex,v 1.2 2004/02/24 11:22:11 spepping Exp $
\usepackage{graphicx}
\usepackage[figuresright]{rotating}

\newcommand{\be}{\begin{equation}}
\newcommand{\ee}{\end{equation}}
\newcommand{\ba}{\begin{eqnarray}}
\newcommand{\ea}{\end{eqnarray}}
\newcommand{\ts}{\textstyle}

\newcommand{\AmS}{{\protect\the\textfont2
  A\kern-.1667em\lower.5ex\hbox{M}\kern-.125emS}}

\title{Model-dependent radiative corrections to $\tau^- \to \pi^-\pi^0\nu$ 
revisited\thanks{Partially supported by Conacyt}}

\author{A. Flores-Tlalpa\address[cinves]{Departamento de F\'\i sica,
Cinvestav,  Apdo. Postal 14-740, M\'exico, D.F., M\'exico},
        F. Flores-Ba\'ez\addressmark[cinves],
        G. L\'opez Castro\addressmark[cinves] and 
        G. Toledo S\'anchez\address{Instituto de F\'\i sica, UNAM, A. P. 
20-364, 01000 M\'exico, D.F., M\'exico}\thanks{Also supported by 
DGAPA-UNAM}}
       
\begin{document}

\begin{abstract}
The long-distance electromagnetic radiative corrections to $\tau^- \to 
\pi^-\pi^0\nu_{\tau}$ are re-evaluated. A meson dominance model is used to 
describe the emission of real photons in this decay. Results obtained for 
the hadronic spectrum and the decay rate in photon inclusive reactions are 
compared with previous calculations based on the chiral resonance theory.
Independent tests in $\tau \to \pi\pi\nu\gamma$ that can help to 
validate the predictions of one of the two models are briefly 
discussed. \vspace{1pc}
\end{abstract}

\maketitle

\section{INTRODUCTION}

Radiative corrections to $\tau^{\pm} \to \pi^{\pm}\pi^0\nu_{\tau}$ 
($\tau_{2\pi}$) decays are important for several reasons:
\begin{itemize}
\item The current precision in the world average of $\tau_{2\pi}$   
measured branching ratios is reaching the 0.4\%  level \cite{pdg06}.
A correct comparison of theory and experiment 
requires the inclusion of $O(\alpha)$ radiative corrections.
\item The conserved vector current (CVC) hypothesis, valid in the isospin 
symmetry limit,  predicts the equality of the weak (measured in 
$\tau_{2\pi}$ decays) and electromagnetic form factors. Measurements 
exhibit departures \cite{Davier06} beyond  expected isospin symmetry 
breaking effects (which include radiative corrections).
\item Predictions of the two-pion vacuum polarization contribution to the 
$\mu$ anomalous magnetic moment based on $\tau$ and $e^+e^-$ data, should 
be equal on the basis of CVC. However, they differ by more than 
$3\sigma$'s \cite{Davier06}. 
\end{itemize}
The CVC hypothesis has been verified with high accuracy (at the 
level of $10^{-4}$) in decay rates of superallowed Fermi transitions 
\cite{Sft:05}. The discrepancies pointed out above suggest that 
unaccounted effects, either in experimental data of pion form factors or 
in isospin  breaking  corrections, may  have escaped consideration.

In this contribution we revisit the long-distance (LD) model-dependent 
radiative  corrections to $\tau_{2\pi}$ decays. LD corrections provide 
an energy-dependent source of isospin breaking correction to be applied to 
the two-pion spectral functions in $\tau$ lepton decays. Previous calculations 
of  the corrections to the hadronic invariant mass spectrum in this decay 
were studied in  refs. \cite{Ciri01,Ciri02} within the framework of 
chiral resonance  theory \cite{Chiral89}. A different approach to 
compute LD 
corrections, based on a meson dominance model, was considered in refs. 
\cite{Nos05,Nos06}. Here we discuss the different results obtained from 
the two approaches and suggest how independent tests can be carried 
out in the corresponding radiative $\tau^- \to \pi^-\pi^0\nu\gamma$ to 
distinguish between the two models.

\section{LONG-DISTANCE CORRECTIONS TO THE HADRONIC SPECTRUM}

The radiative corrected hadronic invariant mass distribution in 
$\tau_{2\pi}$ decays, is obtained by adding the virtual corrections of 
$O(\alpha)$ and the real photon corrections, shown in Figure 1, to the 
zero 
order expression  ($t=(p_{\pi^-}+p_{\pi^0})^2$ is the square of the 
momentum transfer):
\be
\frac{d\Gamma(\tau_{2\pi(\gamma)})}{dt} = \frac{d\Gamma^0}{dt}+ 
\frac{d\Gamma^1_v}{dt}+\frac{d\Gamma^1_r}{dt}\ .
\ee
\begin{figure}[htb]
\vspace{9pt}
\includegraphics[scale=0.43]{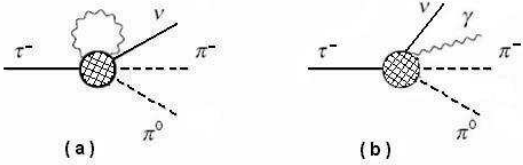}
\caption{Virtual (a) and real (b) photon corrections of $O(\alpha)$ to 
$\tau_{2\pi}$ decays.}
\label{figure1}
\end{figure}
If we also add to the above expression the short-distance corrections 
arising from the emission and reabsorption of gauge and the Higgs bosons, 
we get the fully radiative corrected expression: 
\be
\frac{d\Gamma(\tau_{2\pi(\gamma)})}{dt} =  \frac{d\Gamma^0}{dt} S_{EW} 
G_{EM}(t)\ .
\ee
The factor $S_{EW}= 1.026 \pm 0.0003$ in eq. (2) summarizes the 
short-distance corrections and includes the effects of resummation of 
dominant logarithms to all orders \cite{Sir} and the remaining  
electromagnetic corrections of order $\alpha$ \cite{Bra90}. $S_{EW}$ 
includes also the resummation of sub-leading strong interaction effects 
which were recently discussed in ref. \cite{Erler}.  Given 
that high energy virtual corrections probes the quark level structure in  
semileptonic decays ($\tau^- \to \bar{u}d\nu_{\tau}$), $S_{EW}$ is 
believed to be independent of the specific $\Delta S=0$ $\tau$ lepton 
decay.   

The long-distance radiative corrections are included in the factor 
$G_{EM}(t)$ and are model-dependent. The couplings of photons to hadrons 
are calculated on the basis of scalar QED and also include the effects of 
model-dependent  couplings of the photon to hadrons in all possible ways.
It is defined from eq. (1) as follows:
\ba
G_{EM}(t) &=& 
1+\frac{\textstyle \frac{\ts d\Gamma^1_v}{\ts dt} 
+\frac{\ts d\Gamma^{1,m.i.}_r}{\ts dt} 
+\frac{\ts d\Gamma^{1,m.d.}_r}{\ts dt}}{\textstyle 
\frac{\ts d\Gamma^0}{\ts dt}} 
\nonumber \\
&\equiv & G^0_{EM}(t)+G^{rest}_{EM}(t)\ ,
 \ea
where we have separated the real photon corrections into its 
model-dependent (m.d.) and its model-independent (m.i.) parts (see 
\cite{Ciri02,Nos06} for details). The model-independent correction 
$G^0_{EM}(t)$ includes the sum of virtual corrections and $m.i.$ piece 
of real photon emission necessary to cancel infrared divergences; the 
remaining piece $G^{rest}_{EM}(t)$ is regular and model-dependent.   Let 
us comment that to get the rates for real photon emission in eq. (3) we 
have integrated over all the photon  energies; thus, the $G_{EM}(t)$ 
correction can be applied to photon inclusive $\tau_{2\pi}$ measurements 
only.

  The correction factor $G_{EM}(t)$, was first calculated in 
refs.  \cite{Ciri01,Ciri02}. Both, virtual and real corrections, were 
computed in the 
framework of the chiral resonance theory \cite{Chiral89} by considering 
the exchange of $\rho$ \cite{Pich} and $a_1$ \cite{Kuhn90} resonances.  The 
axial couplings to the weak 
current in real photon corrections were assumed in that model to include 
the axial anomalous terms \cite{wzw}. Figure 2 displays the results 
obtained in ref. \cite{Ciri02} for $G^0_{EM}(t)$ (short-dashed line) and 
$G_{EM}(t)$ (long-dashed line). Notice that the contribution of 
model-dependent corrections $G^{rest}_{EM}(t)$ are small and negative for 
$t\geq 0.5$ GeV$^2$ and positive and rapidly increasing as the 
threshold is approached. 
\begin{figure}[htb]
\vspace{9pt}
\includegraphics[angle=270,scale=0.38]{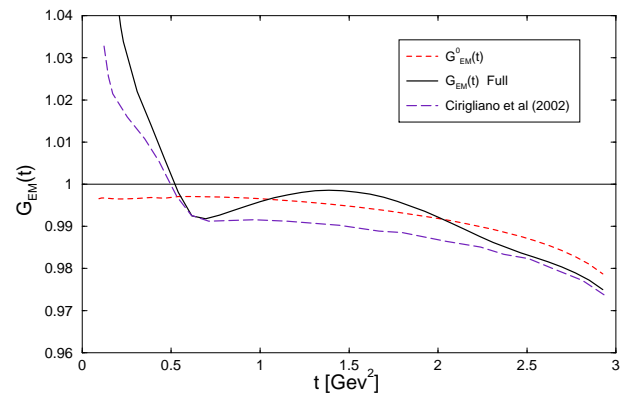}
\caption{Energy dependence of the long-distance correction $G_{EM}(t)$: 
model-independent corrections $G^0_{EM}(t)$ (short-dashed), full 
corrections of ref. \cite{Ciri02} (long-dashed) and ref. \cite{Nos06} 
(solid) are shown.} 
\label{figure2}
\end{figure}

The region of $t$ very close to threshold must be handled with care. 
The apparent divergent behavior in that region arises from the 
kinematical suppression of the tree-level spectrum that appears in the 
denominator of eq. (3). The definition given in eq. (1) must be directly 
used in that case.

In refs. \cite{Nos05,Nos06} the emission of real photons was considered 
in the framework of a meson dominance model. The idea behind our approach 
is that given the large momentum transfer released in $\tau$ decays, all  
intermediate states involving the production and decay of light 
resonances ($\rho(770), \omega(782), a_1(1620)$) (see 
Figure 3) that are allowed by their quantum numbers, must contribute. The 
different couplings entering in the  
model-dependent contributions were determined from independent low 
energy processes (see \cite{Nos05} for details). As we will see below, the 
diagram (g) in Figure 3, which has not been considered in the calculation 
of ref.  \cite{Ciri02}, will  play an important role from a numerical 
point of  view. 
\begin{figure}[htb]
\vspace{9pt}
\includegraphics[scale=0.4]{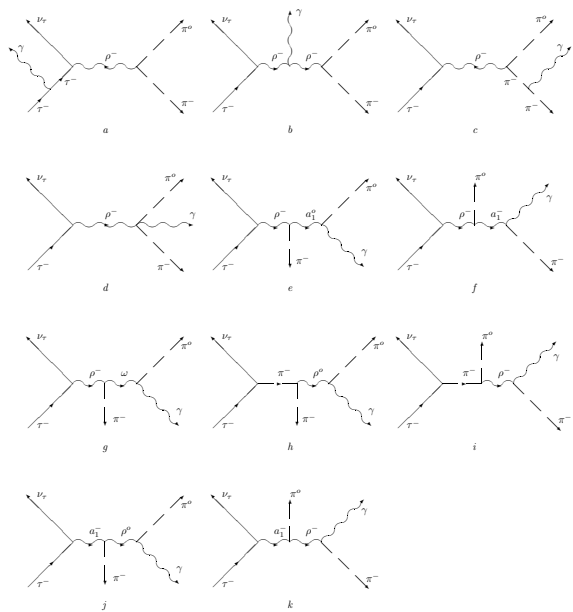}
\caption{Feynman diagram contributions to radiative $\tau \to 
\pi\pi\nu\gamma$ decays. The purely model-dependent contributions are 
shown in diagrams (e-k).}
\label{figure3}
\end{figure}
  In Figure 2 we plot our long-distance correction factor $G_{EM}(t)$ 
(solid line) as a  function of the squared momentum transfer $t$. 
In the region above $t=0.7$ GeV$^2$, our radiative corrections are 
smaller than the ones calculated in ref \cite{Ciri02}. Below that value, 
however, our results are larger. Despite the fact that the predictions of 
both models give small long-distance corrections, the difference in the 
integrated observables, --decay rate and $a_{\mu}^{LO,had}$-- turns out to 
be interesting (see below).

The difference between our calculation of $G_{EM}(t)$ and that of ref. 
\cite{Ciri02} stems almost completely from the anomalous $\rho\omega\pi$ 
vertex (Figure 3g). As it was discussed in ref. \cite{Nos06}, the 
predictions of both models coincide when such diagram is excluded from 
our calculations. This also confirms that the axial-vector contributions, 
despite their very different origin in both models, are almost negligible. 
Since the $\rho\omega\pi$ coupling becomes an important contribution 
within our model, the question arises whether the energy dependence of 
this coupling would affect our predictions in a sizable way. To have an 
idea of the 
answer, we have allowed a variation of $\pm 30\%$ around the central value 
($g_{\rho\omega\pi}=0.o12$ MeV$^{-1}$) used for this coupling in our 
calculations. We have found that a similar variation is obtained, for 
example, in our estimate of the shift in $a_{\mu}^{\pi\pi,LO}$ (eq. 5 
below).

A simple and useful analytical expression can be obtained for the 
long-distance correction factor. In almost all the interval of $t$, 
$G_{EM}(t)$ can be approximated  very well by the polynomial function 
($x=t/m_{\tau}^2$) \cite{Nos06}:
\ba
G_{EM}(x)\!\!\! &=&\!\!\! 1.107-1.326x+5.667x^2 -10.95x^3 \nonumber \\
&& + 9.735x^4-3.2776x^5\ .
\ea

As is well known, experimental data on the $\tau_{2\pi}$ spectral  
function can be used to predict the dominant part of the hadronic vacuum 
polarization contribution to the muon anomalous magnetic moment  
($a_{\mu}^{\pi\pi,LO}$). The model-dependent long-distance corrections 
affect the prediction of ($a_{\mu}^{\pi\pi,LO}$) based on tau data 
(additional sources of isospin breaking corrections were discussed in ref. 
\cite{Davier03}), 
particularly when photon inclusive measurements are used. The correction 
due to LD radiative effects to be applied to the prediction of 
$a_{\mu}^{\pi\pi,LO}$ based on $\tau_{2\pi}$ data can be estimated from 
the  following formula \cite{Ciri02}: 
\ba
\Delta a_{\mu}^{\pi\pi,LO} &=& \frac{1}{4\pi^3} \int dt {\cal K}(t) \left[ 
\frac{K_{\sigma}(t)}{K_{\Gamma}(t)}\frac{d\Gamma_{\pi\pi(\gamma)}}{dt}\right]
\nonumber \\ 
& & \ \ \ \times\left(\frac{1}{G_{EM}(t)}-1\right)\nonumber \\
& =& -3.7 \times 10^{-10} \ . 
\ea
The expression $K_{\sigma}(t)\ (K_{\Gamma}(t))$ in eq.(5) contains the  
kinematical factors and fundamental constants for $2\pi$ production in 
$e^+e^-$ annihilation ($\tau$ decays) \cite{Ciri02} and ${\cal K}(t)$ is 
the kernel function associated to radiative corrections \cite{Ker}. The 
function  
$d\Gamma_{\pi\pi(\gamma})/dt$ must be the measured inclusive photon 
invariant mass distribution (for the purposes of our estimate and 
of comparison with previous calculations we have used eqs. (4.1) and (5.6) 
from ref.  
\cite{Ciri02}).

   The result shown in eq. (5) is almost 4 times larger than the one 
reported in ref. \cite{Ciri02} ($\Delta a_{\mu}^{\pi\pi,LO}=-1.0\times 
10^{-10}$). It is of a size similar to the   
shift in $a_{\mu}^{\pi\pi,LO}(\tau$, based) produced by the 
effects of $\rho-\omega$  mixing  or the pion mass difference in pion 
form factors \cite{Davier03}.

\section{CORRECTIONS TO THE DECAY RATE}

   The corrections to the decay rate can be obtained from 
direct integration of eq. (1). If the emission of hard photons 
($E_{\gamma}\geq \omega_0$) is discriminated  by experiments, it 
becomes useful to define the corrected rate that includes the emission of 
soft photons:
\be
\Gamma(\pi\pi(\gamma),E_{\gamma}\leq \omega_0) = 
\Gamma(\pi\pi)\cdot (1+\delta_{LD})\ ,
\ee
where $\Gamma(\pi\pi)$ is the decay rate without long-distance 
corrections. 

\begin{table}[htb]
\caption{Long-distance corrections $\delta_{LD}$ to the  
integrated rate  of $\tau_{2\pi}$ decays.}
\label{table:1}
\newcommand{\m}{\hphantom{$-$}}
\newcommand{\cc}[1]{\multicolumn{1}{c}{#1}}
\renewcommand{\tabcolsep}{0.73pc} 
\renewcommand{\arraystretch}{1.2} 
\begin{tabular}{@{}lll}
\hline
\cc{$\omega_0$ (MeV)} & \cc{$\delta_{LD}(\%)$} & 
\cc{$\delta_{LD}(\%)$ } 
\\
 & \cc{this work} & Ref. \cite{Ciri02} \\
\hline
$300$                & $-0.31$ & -- \\
$400$     & $-0.27$ & --   \\
$500$     & $-0.23$ & --    \\
$600$     & $-0.19$ & --    \\
$700$     & $-0.16$ & --    \\
$800$     & $-0.15$ & --    \\
$\omega_{max}$     & $-0.15$ & $-0.38$   \\
\hline
\end{tabular}\\[2pt]
\end{table}

   In the second column of Table 1 we show our long-distance corrections 
$\delta_{LD}$ for different values of the cutoff $\omega_0$ for hard 
photons. The correction to the photon inclusive rate corresponds to 
$\omega_0=\omega_{max}$. In this case, our result turns out to be less 
than twice smaller that the correction obtained using the model of ref. 
\cite{Ciri02} shown in the third column (this result was not given in that 
reference; we have  estimated its value by setting $g_{\rho\omega\pi}=0$ 
in our model). 

On the  other hand, we 
observe from Table 1 that the correction to the decay rate due to hard 
photons (let say $E_{\gamma} \geq 300$ MeV) is around $+0.17\%$. This 
result is  much smaller than the estimate ($+0.8\%$) given recently in 
ref. \cite{Ind05} based on the infrared logarithmic term of radiative 
events.

\section{AN INDEPENDENT TEST FOR OUR MODEL}

One may wonder if there is an independent way to discriminate between the 
predictions of meson dominance and chiral resonance models. 
The answer is yes, and radiative  $\tau \to 
\pi\pi\nu\gamma$ ($\tau_{2\pi\gamma}$) decays can be useful for that 
purpose. 
\begin{figure}[htb]
\vspace{9pt}
\includegraphics[angle=270,scale=0.35,]{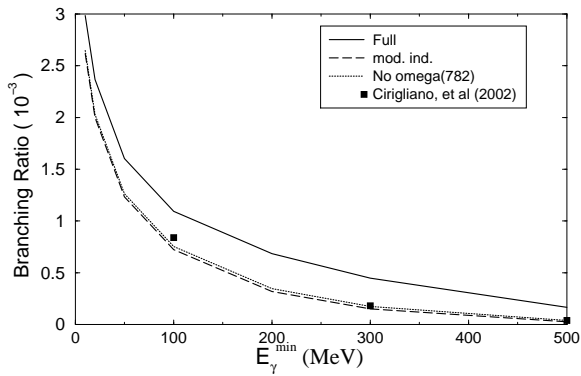}
\caption{Branching ratio of radiative $\tau_{2\pi}$ decay as a function 
of the photon energy cutoff $E_{\gamma}^{min}$.}
\label{figure4}
\end{figure}

In figure 4 we plot the 
branching ratio for this decay as a function of the minimum photon energy 
cutoff (photons of energy larger than $E_{\gamma}^{min}$ are detectable 
in a given experiment). The 
solid line denotes the result of our calculation including all the 
diagrams of Figure 3, while the model-independent (diagrams $a-d$ of 
figure 3) result is represented by the dashed line. Just for comparison, 
we also show (dotted line) the result obtained when the diagram involving 
the $\omega(782)$ meson (Figure 3g) is excluded. The branching ratios 
obtained in ref. \cite{Ciri02} are displayed as three squares at 
$E_{\gamma}^{min}=100, 300$ and $500$ MeV. Clearly, our branching ratios 
differs significantly from the results of 
ref. \cite{Ciri02} for $E_{\gamma}^{min} \geq 150$ MeV and can be a 
useful discriminator between the two models. Other observables associated 
to $\tau_{2\pi\gamma}$ decays that can help to distinguish  the 
predictions of the two models have been discussed in ref. \cite{Nos05}.

In summary, we have compared the long-distance radiative corrections 
obtained in the context of the meson dominance model proposed in refs. 
\cite{Nos05,Nos06} with those obtained in the chiral resonance model 
discussed in refs. \cite{Ciri01,Ciri02}. Despite the very different 
assumptions 
involved in such models, we have found that the only (numerically) 
important difference obtained in the calculation of long-distance 
corrections arise from the real photon emission diagram 
involving the $\rho\omega\pi$ vertex (Figure 3). 

This difference is noticeable in the calculation of the radiative 
correction to the di-pion spectrum and in other observables associated to 
the radiative $\tau$  lepton decay. In particular, long-distance 
corrections shift the two-pion hadronic vacuum polarization contribution 
to $a_{\mu}$  extracted from tau data by $-3.7\times 10^{-10}$, which is 
four times larger than the prediction of ref. \cite{Ciri02}.

Finally, we have pointed out that some observables associated to radiative 
$\tau_{2\pi}$ decays can help to distinguish between the predictions 
of both models.

\end{document}